\documentclass[fleqn,twoside]{article}
\usepackage{espcrc2}
\usepackage{graphicx}

\usepackage{graphicx}
\usepackage[figuresright]{rotating}

\def\gs{\mathrel{\hbox{\rlap{\hbox{\lower4pt\hbox{$\sim$}}}\hbox{$>$}}}}
\def\ls{\mathrel{\hbox{\rlap{\hbox{\lower4pt\hbox{$\sim$}}}\hbox{$<$}}}}
\def\ks{{\rm\thinspace ks}}
\def\km{{\rm\thinspace km}}
\def\ps{{\rm\thinspace s^{-1}}}
\def\kmps{\hbox{$\km\ps\,$}}
\def\et{{et al.\ }}
\def\mcg{{MCG--6-30-15}}
\def\rg{{\thinspace r_{\rm g}}}
\def\ka{{K$\alpha$}}
\def\ovii{{O~\textsc{vii}}}
\def\oviii{{O~\textsc{viii}}}
\def\fei{{Fe~\textsc{i}}}

\def\xmm{{\it XMM-Newton}}
\def\sax{{\it BeppoSAX}}
\def\xte{{\it RXTE}}
\def\asca{{\it ASCA}}

\def\keV{{\rm\thinspace keV}}

\newcommand{\AmS}{{\protect\the\textfont2
  A\kern-.1667em\lower.5ex\hbox{M}\kern-.125emS}}

\title{A long, hard look at \mcg\ with \xmm}

\author{S. Vaughan\address[IoA]{Institute of Astronomy, University
        of Cambridge, Madingley Road, Cambridge CB3 0HA, UK}
        \thanks{Present address: X-Ray and Observational Astronomy
        Group, Department of Physics and Astronomy, University of
        Leicester, Leicester LE1 7RH, UK},
        A. C. Fabian\addressmark[IoA],
         K. Iwasawa\addressmark[IoA] and
         A. K. Turner\addressmark[IoA]}
     
\begin{document}

\begin{abstract}
We summarise the primary results from a 320~ks observation of the
bright Seyfert 1 galaxy \mcg\ with \xmm\ and \sax. 
\vspace{1pc}
\end{abstract}

\maketitle


\section{Introduction}

\mcg\ has received particular attention
since \asca\ observations revealed the presence of a broad, asymmetric
emission feature between $4 - 7 \keV$, identified with a
highly broadened Fe \ka\ emission line  (\cite{T95,I96,I99}).  The
line profile can be explained in terms of fluorescent emission
from an accretion disc extending down to $\ls 6 \rg$
about a black hole ($\rg \equiv GM/c^2$). 
This is the relativistic `disc line' model
(\cite{F89,L91}). The broad Fe \ka\ line therefore potentially offers
a powerful diagnostic of the physical conditions in the immediate
environment of the black hole (see \cite{RN} for a review).

However, modelling the spectrum poses considerable challenges.  Along
with many Seyfert galaxies, \mcg\ shows complex absorption in its X-ray
spectrum.  This complicates the process of identifying the correct
underlying continuum and thereby measuring the superposed line
emission.  Indeed, the presence of relativistically broadened emission
lines in the X-ray spectra of other Seyfert 1 galaxies has recently
become something of a {\it cause celebre} (e.g. \cite{LZ,IM,B01,L01,PR}).

The long ($\sim 320\ks$) \xmm\ observation of \mcg, taken in
July--August 2001 simultaneously with \sax, is  of great importance
since  \mcg\ offers the best established example of a
relativistic line profile.  The results have been discussed in a series of
papers (\cite{F02,FV,V03a,V03b,T03,B03}) and have been compared with the
earlier \xmm\ observation taken in 2000 (\cite{W01}).


\section{Results}

The following are the primary observational
results based on the \xmm\ dataset.

$\bullet$
The two \xmm\ observations (taken in 2000 and 2001) sampled
fairly typical `states' of the source. The former observation 
sampled a period of lower flux than the latter. However, the
long-term \xte\ monitoring shows this can be attributed to short-term
variability and does not imply a secular difference between the
states of the source during the two observations
(\cite{V03b} and references therein). 


\begin{figure}[!t]
\begin{center}
\includegraphics[width=12pc, angle=270]{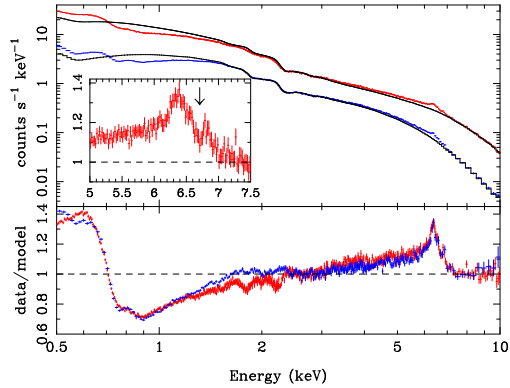}
\end{center}
\vspace{-1 cm}
\caption{
Top: EPIC pn and MOS spectra.
Bottom: Ratio of data to a power-law model joining the $2.5-3 \keV$ data
and $7.5-10 \keV$ data. See \cite{F02,FV}.
}
\label{fig:fit}
\vspace{-0.5 cm}
\end{figure}

$\bullet$
The high energy spectrum obtained from \sax\ shows a strong
Compton-reflection signature (\cite{F02,B03}), 
as did the earlier \xmm/\xte\ observation (\cite{W01}).
There was no evidence for a low energy cut-off or roll-over in the
continuum out to $\sim 100\keV$.

\begin{figure}[!t]
\begin{center}
\includegraphics[width=12pc, angle=270]{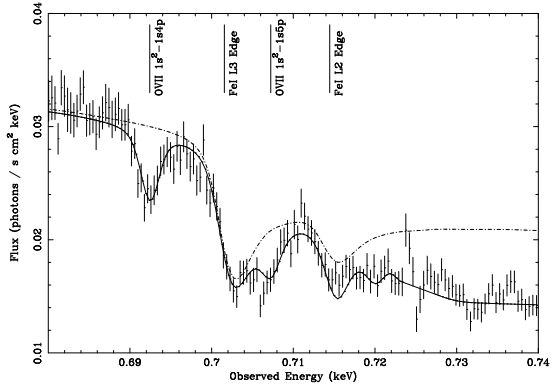}
\end{center}
\vspace{-1cm}
\caption{Close-up of the (fluxed) RGS spectrum 
compared to a model of \ovii, \oviii\ and
\fei\ absorption (solid line) and \fei\ absorption only (dot-dash line).
See \cite{T03}.}
\label{fig:fe_edge}
\end{figure}

$\bullet$
The RGS spectrum shows complex absorption including
\ovii, \oviii\ and \fei\ and a range of other ions
(\cite{S03,T03}; see Fig.~\ref{fig:fe_edge}). The opacity is concentrated
mainly below $\sim 2\keV$ but still has an effect on the
$3-10\keV$ spectrum (section~4.2 of \cite{V03b}).


$\bullet$
Even after accounting for absorption 
the fluorescent iron line is strong and broad (\cite{F02,V03b,W01}). 
The bulk of the line flux is resolved with EPIC. 
The emission peak concentrated around $6.4\keV$ is resolved with a
width $FWHM \sim 4.5\times 10^4$~\kmps, strongly indicating an origin
within $\ls 100\rg$. There is also a significant, asymmetric extension
to lower energies that indicates strong gravitational redshifts.
In addition there is a weak, intrinsically narrow core to the line emission
(\cite{F02,V03b,W01,L02}).

\begin{figure}[!t]
\begin{center}
\includegraphics[width=12pc, angle=270]{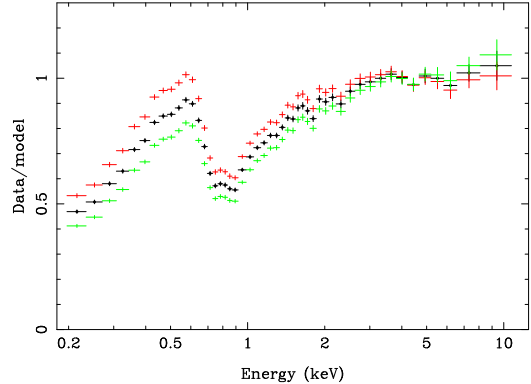}
\end{center}
\vspace{-1cm}
\caption{
EPIC pn difference spectrum produced by subtracting the
low flux spectrum from the high flux spectrum,
shown as a ratio to a power-law (modified 
by Galactic absorption) fitted across the $3-10\keV$ range.
The difference spectrum is consistent with a power-law
component (PLC) modified by warm absorption.
The data/model ratio calculated using the best fitting 
slope is shown with circles. Above and below this lie the
ratios calculated assuming the $90$ per cent lower and upper limits
on the slope of the $3-10\keV$ power-law.
See \cite{V03b}.
}
\label{fig:plc}
\end{figure}

$\bullet$
The $3 - 10\keV$ EPIC spectrum was well fitted using a model
comprising emission from the surface of a relativistic accretion disc.
The strong reflection explains the strength of both the iron line
and the Compton reflection continuum. The best fitting model 
includes emission down to $\approx 1.8\rg$
(\cite{F02,V03b,W01}). 


$\bullet$
Previous observations with \asca\ (\cite{I96,I99,SIF}) and \xte\
(\cite{L00,VE}) showed the 
photon index of the $2-10\keV$ continuum to be correlated with its flux. 
The \xmm\ observations confirm this and demonstrate the trend
is reversed below $\sim 1\keV$, where the spectrum hardens with 
increasing flux (section~5 of \cite{V03b}). The average
variability amplitude is highest in the range $\sim 1 - 2 \keV$, and
is lowest at energies around the iron line. 

\begin{figure}[!t]
\begin{center}
\includegraphics[width=12pc, angle=270]{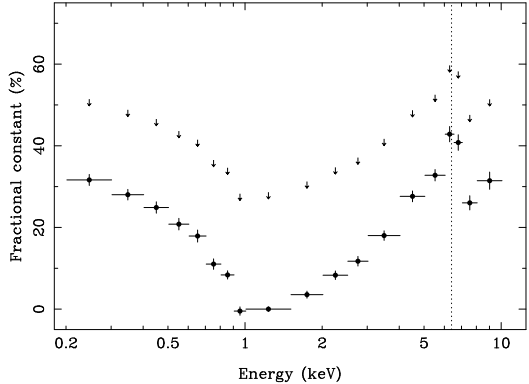}
\end{center}
\vspace{-1cm}
\caption{
Fractional contribution to the spectrum from the constant component
deduced from the linear flux-flux relation. The errors are $1\sigma$  
confidence limits obtained from the linear fit to the flux-flux data.
This spectrum closely resembles a reflection dominated component (RDC).
If the constant component contributes a non-zero fraction in the
$1.0-1.5\keV$ range the contributions at all energies will increase
but the spectral shape will remain the same. The arrows indicate the
absolute upper limits on the fraction of constant emission.
See \cite{V03b}.}
\label{fig:rdc}
\end{figure}

$\bullet$
The variable spectrum can be decomposed into two components, 
a variable Power-law component (PLC: Fig.~\ref{fig:plc}, see \cite{FV,V03b}) and a
relatively constant 
Reflection Dominated Component (RDC: Fig.~\ref{fig:rdc}, see \cite{FV,V03b}). The
spectral variability (at least on timescales $\sim 10\ks$) can be
explained almost entirely by variations in the relative strength of
these two components, caused solely by changes in the normalisation of
the PLC (\cite{FV,V03b,SIF,TUM}). An analysis of the flux-flux plots
(\cite{V03b,TUM}) and the difference
spectra (\cite{FV,V03b}) shows the slope of the PLC remains
approximately constant with flux.

$\bullet$
The EPIC spectrum indicates there is resonance absorption by ionised
Fe at $\approx 6.7\keV$ (\cite{F02,V03b}; Fig.~\ref{fig:fit}). 
This was predicted based on the
presence of soft X-ray warm absorption (\cite{S03,M94}) and
has been observed in at least one other high quality EPIC spectrum
(NGC~3783; \cite{R03}). This resonance absorption appeared to
vary between the two \xmm\ observations
(section~6 of \cite{V03b}).

\begin{figure}[!t]
\begin{center}
\includegraphics[width=12pc, angle=270]{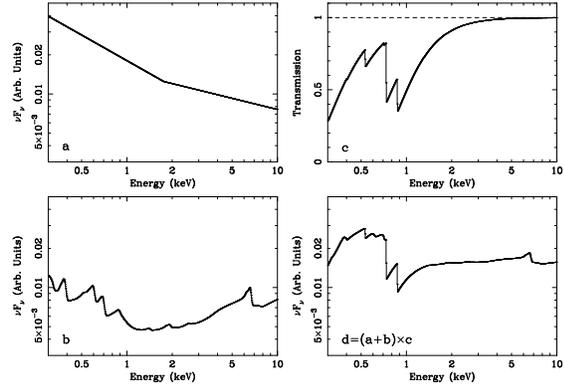}
\end{center}
\vspace{-1cm}
\caption{
Schematic of the components comprising the spectral model.
($a$) Power-law component (PLC).
($b$) Reflection dominated component (RDC).
($c$) Dusty warm absorber (DWA).
($d$) Resulting observed spectrum.
See \cite{T03}.
}
\label{fig:model}
\end{figure}

\begin{figure}[!t]
\begin{center}
\includegraphics[width=12pc, angle=270]{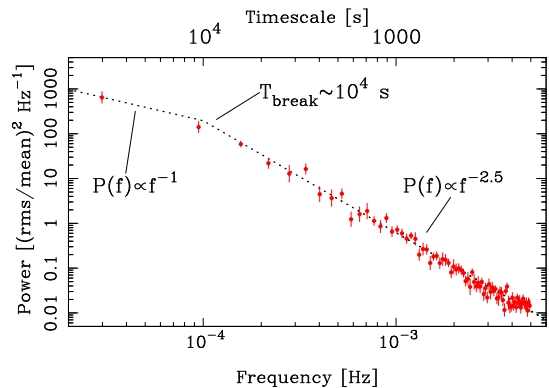}
\end{center}
\vspace{-1cm}
\caption{Power spectral density (PSD) function based
on the $0.2-10\keV$ EPIC pn light curve. The PSD appears
steep at high frequencies and flattens towards lower
frequencies. Shown is the `unfolded' PSD.
See \cite{V03a}.
}
\label{fig:psd}
\end{figure}

$\bullet$
The variations in X-ray luminosity show many striking similarities
with those seen in GBHCs such as Cygnus X-1 (\cite{V03a}; see
Fig.~\ref{fig:psd}).  
In particular, the Power Spectral Density (PSD) function 
is similar to that expected by simply re-scaling the high/soft state 
PSD of Cyg X-1. The continuum variations are energy dependent:
the PSD is a function of energy and the hard variations are
delayed with respect to the soft variations. Similar results
have been found in other Seyfert 1s (e.g. M$^{c}$Hardy's article
in these proceedings).

\section{Conclusions}

A long observation with \xmm\ and \sax\ of \mcg\ in its typical state
has again confirmed the presence of the broad, skewed iron line.   The
best-fitting model includes emission from within  $6r_{\rm g}$,
consistent with emission from a disk around a spinning black hole
(\cite{F02,W01}).

The spectral variability can be explained in a phenomenological
manner. The spectrum comprises the sum of two emission spectra
modified by warm absorption (Fig.~\ref{fig:model}). The two  emission
spectra are a variable Power-Law Component (PLC) and a less variable
Reflection Dominated Component (RDC). The relative constancy of the RDC
causes variations above $\sim 2\keV$ (and particularly  around the
iron line) to be suppressed, and also variations below $\sim 1\keV$
are increasingly suppressed (\cite{F02,FV,V03b}).  Such a
two-component emission model therefore explains the  well known
correlation between $2-10\keV$ spectral slope and flux
(\cite{FV,SIF,VE,TUM}) and the relative lack of iron line  variations
(\cite{V03b,VE}). However, the physics behind the reduced variability
of the RDC is unclear. One intriguing possibility is that gravitational
light bending, which may be strong if the disc extends to $\sim 2\rg$,
is (partially) responsible  (\cite{FV,M03}). 


Based on observations obtained with \xmm, an ESA science mission with
instruments and contributions directly funded by ESA Member States and
the USA (NASA).

\end{document}